\documentclass{article}
\usepackage[utf8]{inputenc}
\usepackage{color}  
\usepackage{latexsym, amscd, amsfonts, eucal, mathrsfs, amsmath, amssymb, amsthm, xr,  makeidx, stmaryrd}
\usepackage{IEEEtrantools}
\usepackage{biblatex}
\title{Substantive general covariance and the Einstein-Klein dispute: A Noetherian approach}
\author{Laurent Freidel$^1$ \and Nicholas Teh$^2$}
\date{%
    $^1$Perimeter Institute for Theoretical Physics\\lfreidel@perimeterinstitute.ca\\%
    $^2$Department of Philosophy, University of Notre Dame\\nteh@nd.edu\\[2ex]%
    \today
}
\begin{document}

\maketitle

\begin{abstract}
    Famously, Klein and Einstein were embroiled in an epistolary dispute over whether General Relativity has any physically meaningful conserved quantities. In this paper, we explore the consequences of Noether's second theorem for this debate, and connect it to Einstein's search for a `substantive' version of general covariance as well as his quest to extend the Principle of Relativity. We will argue that Noether's second theorem provides a clear way to distinguish between theories in which gauge or diffeomorphism symmetry is doing real work in defining charges, as opposed to cases in which this symmetry stems from Kretchmannization. Finally, we comment on the relationship between this Noetherian form of substantive general covariance and the notion of `background independence'.
\end{abstract}

\tableofcontents

\section{Introduction}

In this note, we will explore the enduring impact of Noether's legacy on some of the deepest issues in the philosophical foundations of field theory.
The import of Noether's two theorems is certainly undisputed in both the physics and the philosophy literature, but an articulation of the why and wherefore of her second theorem's import is rather more difficult to come by.
Indeed, most textbooks on field theory only stress Noether's first theorem, and even in those places where Noether's second theorem is mentioned, little more physical insight is offered than the remark that the Noether identities relate the Euler-Lagrange quantities and thus express the `redundancies' that are characteristic of a gauge theory.

The short shrift given to Noether's second theorem goes hand in hand with a glib narrative about a famous epistolary dispute between Einstein and Felix Klein (letters 480 and 492 in \cite{EinsteinVol8}). Contra Einstein, Klein contended that general relativity does not have any physically contentful conserved quantities; the glib narrative takes Noether's second theorem to settle the dispute decisively in favor of Klein. As we will see shortly, the truth is somewhat more complex and to a large extent vindicates Einstein's strategy for responding to Klein. 

A good reason to get clear on the Einstein-Klein dispute is that it holds the key to understanding just how fundamentally Noether's second theorem (albeit not as a piece of mathematics in itself, but as taken up and deployed by physicists) has transformed our understanding of Lagrangian field theory. 
Nonetheless, our reason for considering this controversy is not primarily historical; we wish instead to bear witness to the continuing vitality of Noether's theorems---as a living force that energizes the physical imagination and leads us to make profound new discoveries. 
Our witness consists in using Noether's second theorem to provide a novel and incisive analysis of what \textit{substantive} general covariance really amounts to: the vision that Einstein so desired to contemplate but could only see in a mirror dimly.
While there has of course been much discussion of substantive general covariance---and the related notion of background independence---in the philosophy and physics literature,\footnote{See \cite{pooley2017background, read2016background} and the references therein.} our discussion is (to the best of our knowledge) the first to fully exploit the power of Noether's second theorem in this setting.\footnote{\cite{earman2006two, brown2002general} also discuss Noether's second theorem in relation to general covariance, but our use of the theorem goes far beyond these discussions.}  
Furthermore, we conjecture in Section 3.5 that our Noetherian approach to substantive general covariance provides a royal road to understanding the core phenomenon underlying the elusive notion of `background independence'. 

We begin our reflections in Section 2 in a historico-philosophical fashion: by surveying two questions with which Einstein wrestled long and hard, and which have been re-articulated in various guises in the contemporary philosophy of physics literature.
First, in Section 2.1, we remind the reader of the received view of general covariance (which we call Basic General Covariance) in the context of Lagrangian field theory. Section 2.2 then discusses the question of whether and how general covariance plays a role in extending the Principle of Relativity, and Section 2.3 discusses subject of the Einstein-Klein dispute, viz. whether generally covariant theories have non-trivial conserved charges.
On the basis of the previous two subsections, Section 2.4 formulates what we will call the `Einstein test' for whether a theory has substantive general covariance.

Section 3 develops our novel contribution to the literature: we argue that the machinery of Noether's second theorem provides a powerful technique for determining whether theories have substantive general covariance. 
Crucially (and unlike many other treatements), we do not just make use of the Noether identities, but also of the entire mathematical nexus that relates the Noether identities, the Noether current, and the corner Noether charge which is fixed by a specific choice of Lagrangian. 
Section 3.1 summarizes in an elementary and conceptually transparent way the reformulation of Noether's theorems in the covariant phase space framework (much of which was already anticipated in Noether's original work \cite{noether1918invariante}).
Section 3.2 then applies this machinery to run the Einstein test on several toy theories around which much of the philosophical debate has been centered. 
On the basis of the results of this test, we articulate in Section 3.3 a proposal for what substantive general covariance \`{a} la Einstein amounts to: we give a less demanding form (which only requires a non-trivial corner charge), and a more demanding form (which requires an extension of the Principle of Relativity). 
Section 3.4 revisits the case of General Relativity and the Einstein-Klein dispute from this perspective, and Section 3.5 comments briefly on the relationship between our notion of substantive general covariance and the idea of `background independence': while \citeauthor{pooley2017background} attempts to make the case that the connection between some substantive form of general covariance is `messier, and less illuminating' than the quantum gravity literature might suggest, our analysis suggests that such pessimism is largely a function of considering insufficiently robust versions of substantive general covariance.

\section{What work should `general covariance' do?}

Although the discussion that follows will apply to Lagrangian field theories quite generally (\textit{mutatis mutandis}), we remind the reader that the themes of interest to us were first explicitly problematized in the context of Einstein's General Relativity. Thus, it is here that our meditation will originate. 

The central concept that Einstein struggled so long and hard to formulate, to interpret, and to employ in modeling certain material aspects of the world is that of `general covariance'. 
Today, the descendant of this notion that is commonly invoked by physicists is more plainly referred to as the  `diffeomorphism invariance' of General Relativity, but it is important to appreciate that in the early days of the subject, Einstein's conception of general covariance was very much in flux, both in terms of its mathematical definition, as well as its physical interpretation. 

Here, then, are two key questions that Einstein wrestled with, which are supposed to provide desiderata for a \textit{substantive}---or physically contentful---notion of general covariance:\footnote{Einstein's struggles with general covariance are of course famously various: for instance, we leave aside in this essay the question of the relationship between general covariance and determinism, which has been discussed in the `Hole Problem' literature.}
\begin{enumerate}
    \item Does general covariance play a role in extending the Principle of Relativity?
    \item Do generally covariant theories have non-trivial conserved charges?
\end{enumerate}
As regards the second question, we emphasize that it is clear from Einstein's correspondence with Klein (see letter 480 in \cite{EinsteinVol8} and our discussion in Section 3.4) that the charges Einstein had in mind were \textit{corner} ones, where by `corner' we mean the boundary of a Cauchy surface ---thus a `corner' from the perspective of the entire spacetime manifold.\footnote{Of course, Einstein did not use the terminology of `corners'. What he says is that the construction of the charges is analogous to the is analogous to the use of Gauss' Law in electrostatics, and that the `...possibility of this physically important interpretation is based on the fact that the same quantities [...] which enter in the conservation law [...] can, by virtue of the field equations [...] be written as a ``divergence'','\cite{EinsteinVol8} where by `quantities' he means what we call the Noether charge $J$ below. This means that, returning to standard modern terminology, the on-shell charge integral will descend to a corner by Stokes' theorem.}

Whatever \textit{substantive} general covariance amounted to for Einstein, it was supposed to earn its spurs by means of a `yes' answer to the above two questions. 
We will proceed by laying out in Section 2.1 a `basic' or minimal notion of general covariance that physicists would typically invoke. 
Sections 2.2 and 2.3 will explain Einstein's desiderata for the work that the notion of substantive general covariance needs to do.
Section 2.4 then uses these respective desiderata to formulate a two-part `Einstein test' for substantive general covariance.

\subsection{Basic General Covariance}

According to standard accounts of the history of General Relativity, Einstein initially intended to use general covariance as an adequacy constraint that would single out General Relativity.\footnote{For the history, see the discussion in \cite{norton1993general} and the references therein.} To achieve this goal, he initially proposed that a generally covariant theory is one that can be formulated in such a way that the theory is expressible in any coordinate system (where the different coordinate systems/frames are related by diffeomorphisms).

The problem with this notion of general covariance, as Kretschmann \cite{kretschmann1918physikalischen} pointedly remarked, is that it does not achieve Einstein's goal of singling out a particular theory---just about any theory is generally covariant according to this definition, including Newtonian mechanics, as Cartan showed in his invariant, curved geometric, formulation of Newtonian gravitation (often called Newton-Cartan theory). In light of Kretschmann's response to Einstein, it is now commonly accepted (see \cite{norton2003general} and references therein for discussion) that Einstein's early formulation of general covariance was a misstep. Instead, one should try to provide a `physically contentful' or substantive notion of general covariance that really does pick out some proper subset of physical theories, and perhaps General Relativity in particular. 

Within the philosophical literature, one finds various attempts to articulate a version of substantive general covariance that distinguishes it from the mere freedom to express a theory's fields, equations of motion, and solutions in arbitrary frames (see \cite{pooley2010substantive} for a review).  
For the purposes of this piece, however, it will be helpful to build up to substantive general covariance by starting with the following notion that most physicists take as standard (albeit typically under the head of `diffeomorphism invariance'):\footnote{We will see in Section 3.2 below that only \textit{some} cases of basic general covariance are inequivalent to the freedom to express a theory in an arbitrary frame.} 

\begin{quote}\textbf{(BGC)}
First, distinguish between the background (non-dynamical) and the dynamical fields of the theory. We then say that a Lagrangian of a theory is \textit{basically generally covariant} (BGC) just in case diffeomorphisms of its dynamical fields are variational symmetries\footnote{In other words, diffeomorphisms leave the Lagrangian invariant up to exact terms; we remind the reader that, as a consequence, such diffeomorphisms also take solutions of the theory's equations of motion (EOM) to solutions.} of the Lagrangian.\footnote{For ease of comparison, we note that our \textbf{BGC} is the variational version of what Pooley calls `GC5' in \cite{pooley2010substantive} and what he calls `Diffeomorphism Invariance (final version)' in \cite{pooley2017background}. Pooley also identifies the variational form with what he calls `Background Independence (version 2)' in \cite{pooley2017background}.}\end{quote}

By extension, we will also use \textbf{BGC} below to refer to the analogous case in which diffeomorphisms are replaced by internal gauge symmetries; which case we mean will be obvious from the context. The broad contrast here is between theories whose actions have only rigid symmetries (and thus do not satisfy \textbf{BGC}) and theories with \textbf{BGC}, i.e. whose actions have symmetries that are functions of spacetime (such as diffeomorphism symmetry and internal gauge symmetry). 

It should be clear that \textbf{BGC} is just what physicists typically mean when they say that a Lagrangian theory is diffeomorphism invariant (\textit{mutatis mutandis} for internal gauge symmetry, where the term `gauge-invariant' is used), and it should be equally clear that physicists generally take this feature to be informative about the physical content of the system being modeled.\footnote{As  Norton \cite{norton2003general} says, the feature is taken to physically contentful given an interpretation of the formalism, including which fields in the Lagrangian count as dynamical and which as background.} For reasons that will become clear in Section 3 (and which are amply discussed in \cite{freidel2020edge, freidel2021extended}) we note that we do \textit{not} assume that a theory has a unique action: thus, the Einstein-Hilbert Lagrangian should be understood as just one out of a possible class of standardly stipulated Lagrangians for General Relativity. 

Having put on the table the standard notion of \textbf{BGC}, let us now try to understand what work Einstein would have general covariance do in order to qualify as \textit{substantive}.

\subsection{Extending the Relativity Principle}

We now turn to our first question, which is about the relationship between general covariance and the Principle of Relativity (PR). At one point in the development of General Relativity, Einstein thought that general covariance was simply an extension of the PR in Special Relativity. Later, as part of his response to Kretschmann's critique \cite{kretschmann1918physikalischen}, he retreated to the view that general covariance is necessary but insufficient for extending PR (see \cite{pooley2017background} for a discussion of Einstein's retreat). For our purposes, it will be more fruitful to frame the question thus: does one's preferred notion of general covariance play a distinctive role in extending the Special Relativistic PR (or more generally, analogs of PR for internal gauge symmetry)?
If `yes', then we can consider it to be substantive in virtue of playing this role. 

Here again one finds the familiar story of Einstein making a bold and potentially physically insightful conjecture, but with murky conceptual foundations, and which was thus amenable to being problematized by others.
Correspondingly, there has been much controversy about whether \textit{any} notion of general covariance can play a role in extending PR.
At least within the philosophical literature, one finds various authors (e.g. \cite{belot2000geometry} and \cite{norton1993general}) who answer `no' because they claim that the desired extension of the PR to General Relativity must be trivial, in the sense that the resulting `relativity' group is just the identity. 
In brief, these authors argue for their claim by appealing to the following controversial assumption about PR, which we will call \textbf{(Geometric PR)}: in such an extension, the empirically significant group of symmetries invoked in the Principle of Relativity should be identified with the stabilizer group of the spacetime metric in General Relativity (just as it is typically identified with the stabilizer group of the Minkowski metric in Special Relativity).

If one accepts \textbf{Geometric PR}, then since a generic metric---which is now a dynamical field---in General Relativity has no non-trivial automorphisms, it immediately follows that the PR is trivial in General Relativity. Thus, no form of general covariance (including BGC) can play a role in enabling such an extension. We stress that this conclusion is wholly driven by the assumption of Geometric PR: if the argument works, it works regardless of what one takes substantive general covariance to be.

The reason \textbf{Geometric PR} is controversial is that it is far from clear that this is the most physically perspicuous and relevant interpretation of the PR. Indeed, in \cite{brown1995meaning}, Brown and Sypel articulate a version of the PR which they take to be both the more physically perspicuous interpretation, as well as the one that more faithfully reflects Einstein's own understanding of PR. According to Brown and Sypel, the symmetry group of PR is to be understood as the symmetry group relating inertial frames, which are in turn to be fleshed out as---in idealization---ways of encoding the equivalence of outcomes of experiments set up with the same initial conditions in an \textit{isolated} subsystem. Thus, when in Special Relativity one goes on to define a geometric object (the Minkowski spacetime metric) whose stabilizer group is precisely the symmetry group of the PR, one is merely codifying `...aspects of the comparative behaviour of different systems of physical rods and clocks in relative motion', where the behavior of the target subsystem (whose inertial frames are related by these symmetries) is being measured with respect to an environment frame, from which the subsystem is dynamically isolated. Of course, once one understands this point, one can abstract away from a concrete material system and even its idealization in terms of inertial frames, and arrive at a kind of `iconic' or `generic' image that we call `Minkowski spacetime', which provides an efficacious vehicle for abstract reasoning. This generic `type' can then be made to descend once again into a representation of a concrete empirical scenario when it is filled in with the description of the relevant subsystem-environment decomposition, initial and boundary conditions including the relevant conditions for `isolation', and the practical knowledge of the experimentalist.

Let us call the understanding of PR that we have just articulated \textbf{Embodied PR}, because of its emphasis on how the Principle of Relativity (and its symmetries) needs to be understood not just formally, but as an embodiment of concrete empirical scenarios and their idealizations (such as boundary conditions). It is this embodied understanding of PR (and of symmetries more generally) that we will adopt below, because we are interested in the physical and empirical content of symmetries, which stems from an understanding of how the formal apparatus is used to represent particular subsystem-environment relationships and the related boundary conditions.

Notice that, according to \textbf{Embodied PR}, one needs to be able to understand how a theory models an `isolated subsystem' \textit{before} one can even articulate PR within that theory. Thus, advocates of Embodied PR will not accept the above argument for Geometric PR, because it is based on a definition of PR that ignores the question of how `dynamical isolation' is to be modeled in the extended scenario.\footnote{Ironically, Brown \cite{brading2004gauge} has expressed sympathy for Geometric PR on the grounds that the diffeomorphism symmetry of GR `...does not have an active interpretation in terms of isolated subsystems of the universe.' The debate about whether and in what sense local symmetry (of which the diffeomorphism symmetry of GR is an instance) can have such an interpretation has by now played out quite fully (see \cite{teh2016galileo} and references therein), but our interest here is in highlighting that, on an embodied view (and again: in contrast to Geometric PR), one's conception of substantive general covariance has to do real work in adjudicating whether (Non-extension) is true.} Furthermore, a moment's thought will show that `dynamical isolation' is a much more sophisticated notion in General Relativity than it is in Special Relativity: we typically model it as the asymptotic flatness of a subsystem spacetime. 

To recapitulate, we have arrived at the following way of sharpening our second question: is there a notion of general covariance that has a distinctive role to play in extending PR---in the sense of Embodied PR---from Special Relativity to General Relativity? If we frame things from Einstein's perspective, then the question becomes: can we articulate a suitably generalized version of PR if we are concerned with asymptotically flat subsystems? But it is important to note that the question can be generalized so that we consider isolated finite subsystems as well, and not just asymptotic infinity.\footnote{Furthermore (although it lies beyond the scope of this paper), it is also physically relevant to consider non-isolated subsystems in which charges are not conserved because of symmetry-breaking.}

In Section 3, we will see that Noether's theorems strongly suggest that the answer to the above question is `yes', but before that we will need to turn to a second related facet of Einstein's struggle with general covariance, and the one that led Noether to write her 1918 paper. 

\subsection{Non-trivial corner charges}

The second question of this section is: is there a notion of general covariance that yields physically meaningful corner charges (such as energy)? For instance, General Relativity is a theory whose Lagrangian has the diffeomorphism symmetries of BGC: does it in virtue of this have physically meaningful charges? (We will see in Section 3 that, from the perspective of the framework developed by Noether, it is clear that this question is intimately related to our first question.) 


Historically speaking, this question found its genesis in Hilbert's variational treatment of General Relativity (see \cite{rowe2019emmy, rowe2021emmy} for a masterful treatment of the history, which we follow here). In addition to deriving the field equations in his treatment, Hilbert's main contribution (which he called `the most important goal of his theory' ) was the formulation of an invariant `energy vector' and a proof of its conservation. In response, Einstein provided his own somewhat different variational approach to General Relativity; in this note he derived an energy current which can be written in modern notation as $J_X = C_X + dU_X$, where $C_X$ is the constraint associated with the gauge transformation $X$. $C_X$  vanishes on-shell. $U_X$ is referred to as the charge aspect  or  superpotential (these formulae are generic for a gauge theory, but we will provide explicit details for electromagnetism in the next section).\footnote{We direct the reader to \cite{de2021noether}, forthcoming in this volume, for an explication of Einstein's notation $J = \mathfrak{T} + \mathfrak{t}$ and how he thought about the objects $\mathfrak{T}$ and $\mathfrak{t}$.} Einstein also stressed the importance of the conservation law $dJ_X=0$ for the on-shell current.

It was this series of developments that set the stage for Emmy Noether's contribution to the mathematical foundations of General Relativity. First, as Felix Klein's assistant, Noether worked out the exact relationship between Hilbert's energy vector and Einstein's energy current. 
This analysis, in conjunction with Klein's own work on the topic, led to a open letter from Klein to Hilbert in which Klein explained that the energy current $J_X$ could always be decomposed as the sum of two parts, the first of which vanishes on-shell, and the second of whose divergence is identically zero (where we emphasize that the latter is an off-shell statement). In consequence, Klein asserted, the statement that $dJ_X=0$ is merely a mathematical identity and does not have any physical content.

Upon reading this letter, Einstein wrote Klein to express admiration for his insights, but also to protest that `I regard what you remark about my formulation of the conservation laws as incorrect.' According to Einstein, the conservation of the current was not itself a mathematical identity, but was instead the consequence of a mathematical identity \textit{and} the equations of motion. Einstein further argued that the physical interpretation of such a conservation statement was analogous to the integral form of Gauss's law in the following sense: since the current $J_X$ can be shown to be on-shell exact, i.e. $J_X\approx dU_X$, one can use Stokes' theorem to write the on-shell charge (the integrated current) as a \textit{corner} (relative to the spacetime manifold) quantity, i.e. $\int_\Sigma dU_X = \int_{\partial \Sigma} U_X$. He then proceeded to sketch a particular subsystem for whose boundary conditions (essentially a primitive version of asymptotic flatness) such a computation of charge made sense. 

As regards Einstein and Klein, their further exchange only led to a stalemate, with Einstein insisting that his conception of energy conservation in General Relativity was physically contentful, and Klein denying this claim on the grounds that `mathematical identities' cannot be physically contentful statements.
But against the background of this impasse concerning physical interpretation, mathematical progress was made: Noether published her 1918 `Invariante Variationsprobleme' \cite{noether1918invariante}. 
Herein, she proved two theorems (referred to in the physics community as Noether's First Theorem and Noether's Second Theorem respectively) and derived a corollary (the so-called `Hilbertian assertion') that bore directly on the Einstein-Klein debate. In our terminology, the corollary states that Lagrangian field theories with \textbf{BGC} only have (in virtue of that \textbf{BGC}) trivial currents, where we remind the reader that a `trivial current' is one that can be written as the sum of an exact form and a term that vanishes on-shell (i.e. our $C_X$ and $dU_X$ above respectively). Thus, the on-shell current is identically conserved and is said to have an `improper conservation law'.

It would be a mistake to read Noether here as weighing in on the question of physical interpretation; rather, one should understand her as offering a mathematician's conceptual analysis: given that you think the concepts of BGC and trivial currents are physically relevant, you should be aware of the following mathematical relationship between them.
Read this way, Noether's corollary does not answer our second question, but it does prompt the following specific version of it: `Does a theory with BGC have---on that basis---physically meaningful conserved charges?' where we should now understand the latter as `physically meaningful charges that stem from trivial currents', since (as Noether showed) non-trivial currents cannot be derived from the diffeomorphism (or gauge) symmetry of BGC. 

\subsection{The Einstein test for `substantive'}
To sum up this section, reflection on Einstein's struggles with `extending the principle of relativity' and `defining conserved charges' within the context of GR leads us to what we might call the `Einstein test' for a \textit{substantive} notion of general covariance: 
\begin{enumerate}
    \item The `corner charge' part of the Einstein test (corresponding to Section 2.3): Does the notion of substantive general covariance yield non-trivial corner charges?\footnote{We note that, based on the history, one might reasonably take the corner charge part of the test to include the \textit{conservation} of non-trivial corner charges. However, for the purposes of analytical clarity, it will be convenient for us to make the minimum criterion the existence of \textit{non-trivial} corner charges, with conservation left as a further criterion.}  
    
    \item The `Extending PR' part of the Einstein test (corresponding to Section 2.2): Does the notion of substantive general covariance allow us to extend the principle of relativity (in the sense of Embodied PR that we discussed above)?
    
\end{enumerate}

Given the status of BGC as the `received' version of general covariance (when formulated variationally) and Einstein, Klein and Noether's own recognition of the relevance of BGC to the question of conserved charges, it is natural for us to first try to run the Einstein test on BGC. 
In the next section, we will see how the raw mathematical materials of Noether's theorems can taken up---in the art of the physicist---to give a particularly trenchant formulation of this test: we will find that BGC does not pass the test, but also that the machinery of Noether's second theorem suggests a version of substantive general covariance that does. 

\section{General covariance through a Noetherian lens}

Since Kasia Rejzner's essay \cite{rejzner2020bv}  in this volume expounds on Noether's two theorems from a more technical perspective, namely the BV complex, we shall here take a more down-to-earth approach in which we work in ghost degree zero. However, we stress that the fundamental observations that we make can be reconstructed in terms of the cohomology of the BV complex---a beautiful connection with another part of Noether's career since she was one of the developers of algebraic homology groups, albeit not in the context of her two theorems. 
To set the stage, we note that at a highly schematic level, the chief moral of Noether's two theorems for Lagrangian field theory is that the space of such schematic theories can be partitioned according to whether or not the theories possess two features, which in turn correspond to the symmetry-structure of the theories' Lagrangians. These two features are:
\begin{itemize}
    \item Noether currents. A \textit{non-trivial} Noether current is one that cannot be written as the sum of an exact form and a term that vanishes on-shell. A \textit{trivial} Noether current is one that can be so written.
    \item Noether identities, viz. relations between the Euler-Lagrange quantities that take the form $N^i [\frac{\delta L}{\delta \phi^i}]=0$; we emphasize that since these are mathematical identities, they hold off-shell. A \textit{nontrivial} Noether identity is one whose coefficients $N^i$ do not vanish on-shell. A \textit{trivial} Noether identity is one whose coefficents vanish on-shell.

\end{itemize}
Furthermore, the correspondence to the symmetry-structure is given by Noether's two theorems (and their modern interpretations), as follows:\footnote{Here our presentation of the `iff' statement is heuristic. Making this an honest theorem requires considerable care about how to define equivalence classes of symmetries and currents, as demonstrated by Olver in \cite{olver2000applications}.}
\begin{itemize}
\item Rigid symmetries (symmetries with a finite set of parameters) are in 1-1 correspondence with non-trivial Noether currents. Furthermore, these currents are conserved when we assume the on-shell property in computing the divergence of the current. 

\item Gauge symmetries (symmetries with an infinite set of parameters, i.e. symmetries in the sense of \textbf{BGC}) are in 1-1 correspondence with non-trivial Noether identities. Furthermore, they yield trivial currents whose on-shell expression is conserved as a mathematical identity, i.e. without using the on-shell property in computing the divergence.
\end{itemize}

Let us now descend from this schematic discussion to a more realistic physical context in which we consider boundary conditions of subsystems. In particular, we will use the covariant phase space framework to understand how Noether's theorems get applied in such a context. 

\subsection{The covariant phase space approach to Noether}

First, we sketch the spacetime geometry of the subsystems that we will consider. It will suffice to consider the elementary case of a (simply connected) spacetime $M$ with timelike boundary $\partial M$, and into which we insert an initial Cauchy surface $\Sigma$; the corner $S=\partial \Sigma$ lies in $\partial M$. Introducing such a boundary $\partial M$ is crucial for modeling the boundary conditions of a subsystem; furthermore, one can think of such a finite boundary model as a regularized version of a subsystem with asymptotic boundary conditions. 

Consider the equivalence class $[L]$ of Lagrangians on $M$, where the equivalence relation is given by $L \sim L + d\ell$, i.e. two representatives of the class differ by a boundary Lagrangian $\ell$ (on $\partial M$). 
Such an equivalence class corresponds to a set of bulk equations of motion. 
In the literature, there are two ways of interpreting the representatives of this class, corresponding to two ways of interpreting the fundamental variational formula, a version of which appeared in Noether's 1918 paper, i.e. 
\begin{equation}\label{var}
    \delta L = - \mathcal{E} \delta \phi + d \theta,
\end{equation}
where we remind the reader that $\mathcal{E}=E_{\phi} \delta \phi$ is a field space $1$-form whose coefficient $E_\phi$ is the Euler-Lagrange quantities, and $\theta$ is a field space $1$-form called the (pre-)symplectic potential current. 

According to the first approach, $\theta$ is defined by the fundamental variational formula and the representatives correspond to an ambiguity in the definition of $\theta$.
According to the second approach (for which see \cite{freidel2021extended, de2021noether} and references therein), the variations are only well-defined when we know the full bulk and boundary data---corresponding to a particular representative $L$ (including a possible boundary Lagrangian piece)---and this fully specified data picks out a unique $\theta$ by means of Anderson's homotopy operators \cite{anderson1989variational}; this unique $\theta$ is then the one that appears in the fundamental variational formula.\footnote{We also assume that we have fixed a particular set of coordinates in field space.} 
Here we will follow the second approach, because the incorporation of the boundary data is crucial to resolving the questions we have raised in Section 2. 

We remind the reader that given the unique $\theta$ corresponding to some $L$, we can construct the presymplectic form $\Omega := \int_\Sigma \delta \theta$.
If the Lagrangian has a symmetry $\xi$, we can also construct a corresponding Noether current $J_\xi := I_\xi \theta$ and Noether charge $Q:=\int_\Sigma J_\xi$, $\xi$ is the symmetry vector field on field space and $I$ is the field space interior product.\footnote{We note that, below, we will adopt a standard abuse of notation in which the notation for a vector field on field space $\xi$ is also used to represent the gauge parameter that appears in the co-efficient of that vector field.}
The statement of Noether's first and second theorem can then be understood in terms of these quantities. 

From this perspective, the basic statement of Noether's first theorem arises when $\xi$ is a variational symmetry of $L$, i.e. $L_\xi L = dl_\xi$ (where $L_\xi$ is the field space Lie derivative). We can then use Cartan's magic formula and (\ref{var}) to obtain the conservation statement $dJ_\xi = I_\xi \mathcal{E}$, where the Noether current $J_\xi$ is defined as $J_\xi := I_\xi \theta - l_\xi$.
Furthermore, when $\xi$ is a rigid symmetry then $J_\xi$ is a non-trivial Noether current, in the sense discussed above, and so the current is conserved only on-shell.  

On the other hand, when the symmetry $\xi$ is a non-constant function of spacetime and thus has an infinite number of parameters, we find that we can write $I_\xi \mathcal{E} = d C_\xi + \mathcal{N}_\xi $ where $C_\xi$ is a constraint term that vanishes on-shell, and $\mathcal{N}_\xi$ is the sum of terms linear in $\xi$, each of which contains a differential operator acting on the Euler-Lagrange quantities. We can then use the locality of $\xi$ to argue that $\mathcal{N}$ must vanish as an identity---this yields the non-trivial Noether identity (often also called a Bianchi identity) corresponding to the gauge symmetry $\xi$.
By combining this observation with $L_\xi L = dl_\xi$ and (\ref{var}), we can conclude that $d (J_\xi - C_\xi)=0$ and so the Noether current takes the form $J_\xi = C_\xi + dU_\xi$,
where the `superpotential' or `charge aspect' $U_\xi$ depends on the gauge parameter $\xi$ in a linear but non-trivial manner. In other words, we have just shown that gauge symmetry corresponds to a trivial Noether current, in the above sense.
There are two extremely important points to emphasize about the superpotential $U_\xi$ at this stage:
\begin{itemize}
    \item The superpotential $U_\xi$ is uniquely determined by our choice of Lagrangian in the equivalence class $[L]$ and the symmetry generator $\xi$.\footnote{See \cite{freidel2021extended} for discussion and see \cite{de2021noether} for an explicit formula for $U_\xi$.}
    
    \item For gauge theories, i.e. theories with \textbf{BGC}, the Noether charge $Q^{\Sigma}_\xi:=\int_\Sigma J_\xi = \int_{\partial \Sigma} U_\xi$ is always a \textit{corner} quantity, because the Noether current is on-shell exact, and  determined by the superpotential $U_\xi$. 

\end{itemize}

With these observations in hand, we can see that it is the `corner charge' part of the Einstein test which is more fundamental than the `Extending PR' part of the test: if the corner charges for a gauge theory are trivial (i.e. vanishing), then there is no sense in which the PR can be extended (from a rigid precursor theory) to that gauge theory. If, on the other hand, the corner charges are non-trivial, then subject to understanding what kinds of boundary conditions model an `isolated system', it may be possible to obtain a generalized version of PR in that gauge theory.


\subsection{Test cases}
With the above Noetherian machinery in place, we are now in a position to run the `corner charge' part of the Einstein test on BGC, i.e. to pose the question: does a theory with BGC thereby have non-trivial corner charges? To that end, it will be helpful to make a connection here with some of the toy models that Pooley \cite{pooley2010substantive, pooley2017background} and others use to test the validity of different definitions of general covariance, although we warn the reader that for reasons of simplicity, we have sometimes opted for a different labeling convention from that of Pooley \cite{pooley2017background}.

Consider first the following theory of a free scalar field on Minkowski space that we will call \textbf{SR1}:\footnote{Pooley \cite{pooley2017background} calls the non-variational formulation \textbf{SR1}.}
\begin{equation}
L (\phi, \eta)  = - \frac{1}{2} \eta^{ab} \partial_a \phi \partial_b \phi,
\end{equation}
where the scalar field $\phi$ is the dynamical variable and the Minkowski metric $\eta$ is a non-dynamical (or background) structure. Our variational formula for this theory is $\delta L = E_{\phi} \delta \phi + \nabla_a \theta^a$, where the Euler-Lagrange expression is $E_\phi = \square \phi$ and the pre-symplectic potential current is $\theta^a = - \nabla^a \phi \delta \phi$.
We note that the conservation of the stress-energy tensor $T^{ab}$ follows from applying Noether's first theorem to the rigid spacetime symmetries of this theory. 

Clearly, this theory does not satisfy \textbf{BGC}, because diffeomorphisms are not a variational symmetry of $\phi$.
In our next example, we will modify \textbf{SR1} slightly to show that one can produce an example of a theory that satisfies BGC, but which fails the `corner charge' part of the Einstein test, because it only has a trivial superpotential, and thus the corner charge vanishes. 

\subsubsection{BGC with trivial superpotential}
Consider now a theory that we will call \textbf{SR2} which is given by,
\begin{equation}
    L^{\text{\textbf{SR2}}}(\phi, Y ; \eta) = L(\phi, Y^* \eta), 
\end{equation}
where $Y: \mathbb{R}^d \rightarrow M$ is a parametrization field that captures our freedom to choose coordinates on the spacetime $M$.\footnote{Pooley \cite{pooley2017background} calls this \textbf{SR5}, and his example of \textbf{SR4} is similar.} Due to the latter interpretation of $Y$, it is appropriate to call \textbf{SR2} a `Kretchmannized' version of \textbf{SR1}.\footnote{We note in passing that Yang-Mills theory and General Relativity can be analogously Kretchmannized, and this leads to the construction in \cite{Donnelly_2016}.} Our dynamical variables in this theory will be $\phi$ and $Y$, whereas the Minkowski metric $\eta$ still retains its status as a non-dynamical background structure.

We can again compute the variational formula (\ref{var}) for this theory, thus obtaining $\delta  L^{\text{\textbf{SR2}}} = E_\phi \delta \phi + E_a \chi^a + \nabla_a \theta^a$, where $\chi := \delta Y \circ Y^{-1}$, $E_a = -2 \nabla_b T^{ba}$, and $\theta^a = 2T^{ab} \chi_b - \partial^a \phi \delta \phi$. Notice that in addition to the previous equation of motion $E_\phi = 0$, we now have a new equation of motion $E_a=0$, i.e. the conservation equation for the stress-energy tensor. Furthermore, by inspecting $\theta$, we can see that in addition to the original symplectic pair $\phi$ and its momentum $\nabla^a \phi$, we now have a new symplectic pair $Y^a$ and its momentum $T^{ab}$.

We now draw the reader's attention to an interesting point: clearly, $L^{\text{\textbf{SR2}}}$ is covariant in the sense of BGC (this is true even though $\eta$ remains a non-dynamical field and is not varied) and thus this theory must have a non-trivial Noether identity.
By turning the crank of the Noether machine, we find that this identity is $E_\phi \partial_a \phi - 2 \nabla_b T^{ba}=0$ (that this is an identity can be straightforwardly checked from the definition of $T_{ab}$). 
In other words, the content of the Noether identity \textbf{SR2} is actually the content of Noether's first theorem for \textbf{SR1}---one might call this the \textit{transmutation} of Noether's first theorem into Noether's second theorem via Kretchmannization!

It is also interesting to compute the (trivial) Noether current, i.e $J^{a}_{\xi} = I_\xi \theta -\iota_\xi L^{\text{\textbf{SR2}}} = 0$. In other words, the current vanishes identically off-shell, and we can deduce that the superpotential $U_\xi$---and thus the corner charge algebra---is trivial.

To sum up, the theory \textbf{SR2} demonstrates why BGC fails the `corner charge' part of the Einstein test for substantive general covariance (and in consequence fails the `extending PR' part of the test): even though this theory clearly has BGC, it nonetheless fails to deliver non-trivial corner charges.

\subsubsection{BGC with non-trivial superpotential}

Next, we turn to a familiar example in which BGC does yield a non-trivial corner charge (and thus passes the `corner charge' part of the Einstein test). Consider the Maxwell gauge theory Lagrangian
\begin{equation}
    L= -\frac{1}{4} F \wedge \star F,
\end{equation}
where we have chosen a representative $L \in [L]$ with vanishing boundary Lagrangian, i.e. $\ell=0$.
The variational formula is
\begin{equation}
\delta L = - E ~ \delta A + d\theta,
\end{equation}
where $\delta$ is the exterior derivative on the space of fields, $E= d \star F$ is the Euler-Lagrange expression, and $\theta = \star F ~ \delta A$ is the presymplectic potential current. 

In this case, the gauge symmetry vector field on field space has the form $\hat{X} = dX(x,t) \delta/ \delta A$, and $L_{\hat{X}}$ is the field space Lie derivative with respect to this vector field, although we will now abuse notation by using $X$ to denote both the field space vector field and the local gauge parameter. Since $L_{X} L = 0$, we can repeat the general analysis summarized in Section 3.1 to obtain 
\begin{equation}
    0 = L_{X} L = d(I_{X} \theta) - I_{X} (E \delta A)
\end{equation}
and then by using integration by parts and the definition of the current $J_X = I_{X} \theta$, we have
\begin{equation}
    d ( J_X - EX) = - X dE.
\end{equation}
Since $X$ is an arbitrary gauge symmetry, we can integrate both sides over a domain with boundary, apply Stokes' theorem to the LHS, and---by assuming that $X$ and its derivatives vanish on the boundary---deduce that $dE=0$ as an identity. This is a simple example of a non-trivial Noether identity (here the Noether operator $N$ is just the exterior derivative $d$, which does not vanish on-shell). We note that  $dE= dd*F = 0$ is clearly a mathematical identity (the Bianchi identity) since it follows from $d^2=0$. 

By applying the Noether identity $dE=0$ and the Poincare lemma, we see from (7) that 
 \begin{equation}
    J_X = EX + dU_X \approx dU_X,
\end{equation}
where `$\approx$' denotes `on-shell equality'.\footnote{This is nothing other than the $U(1)$ gauge theory version of Einstein's formula $J = \mathfrak{T} + \mathfrak{t}$, of which the reader can find an analysis in \cite{de2021noether}.
 It explains why energy pseudotensors in GR are often said to be defined only up to a superpotential.}
In other words, we see that the current $J_X$ that is associated with the gauge symmetry is trivial, although---as we are about to see---this of course does not imply that the Noether charge is trivial!

Given that we have a current $J$, we can define a charge by integrating $J$ over a spacelike Cauchy surface $\Sigma$ to obtain:
\begin{equation}
    Q^\Sigma = \int_{\Sigma} J
\end{equation}
We now remind the reader that a unique possibility arises for \textit{trivial} currents (but not non-trivial currents), which is a direct consequence of the \textit{non-trivial} Noether identity: by Stokes' theorem, a trivial on-shell current $J_X \approx dU$ can be converted into a purely boundary Noether charge

\begin{equation}
    Q^\Sigma_{X} = \int_{\Sigma} dU_X \approx \int_{\partial \Sigma} U_X =  \int_{\partial \Sigma} X (\star F),
\end{equation}
where in the last equality we used the explicit form of the superpotential $U_X$ (fixed by $X$ and our choice of $L$), which---unlike the case of \textbf{SR2}---is non-trivial and leads to a non-trivial corner charge algebra. 

To go further and say more about the conservation of this corner charge, one needs to impose boundary conditions on the subsystem that rule out the leakage of `symplectic flux' from the system \cite{harlow2020covariant, freidel2021extended}.\footnote{In general, one also needs to make sure that the chosen boundary configuration has stabilizers, although this condition is guaranteed in our example, since all $U(1)$ gauge field configurations have stabilizers (the constant gauge transformations).}
In the context of the present example, consider the case in which the subsystem has Dirichlet boundary conditions, i.e. $\delta A|_{\partial M} = 0$. Notice that this sets the symplectic potential to zero at the boundary, which prevents symplectic flux leakage.
Then, the symmetries of the subsystem need to preserve these boundary conditions, and so $d X|_{\partial M} = 0$. Evidently, when $X|_{\partial M} = 0$, the symmetry is in the kernel of $\Omega$ and so does not represent a physical symmetry generated by a charge, leaving non-zero constant transformations at the boundary $X$ as the only remaining possibility (which are precisely the non-trivial stabilizers of the boundary condition).

In the above example, we see that the Noether charges are also the canonical conserved charges associated with our boundary condition of interest, namely the Dirichlet boundary condition. The latter are also called `Hamiltonian charges' because the charges are Hamiltonians for the corresponding symmetry transformations in phase space. We note that the Dirichlet boundary conditions represent an \textit{isolated} subsystem, for which one expects the relevant Noether/Hamiltonian charge to be conserved, unlike an open system which is continually experiencing an exchange of charge with its environment.

However, as we will soon see, there can be more complex scenarios in which a Hamiltonian charge associated with a particular boundary condition is the Noether charge of $L + d\ell$, but not of $L$. In Section 3.4, we will discuss how one can resolve this tension between Noether and Hamiltonian charges, which is relevant to the `extending PR' part of the Einstein test of substantive general covariance.

\subsection{Substantive general covariance}

We have just seen that, due to examples such as \textbf{SR2}, BGC does not on the whole pass the `corner charge' part of the Einstein test for substantive general covariance. On the other hand, we have also seen that central examples of theories with BGC (electromagnetism, and by a simple extension: Yang-Mills theory and General Relativity) do yield non-trivial corner charges.
The general moral is that while BGC is insufficient for the existence of non-trivial corner charges, it is nonetheless necessary, because one needs a trivial Noether current in order to obtain a \textit{corner} charge.
What then needs to be added to BGC to obtain a non-trivial corner charge? 

As it turns out, this is one of the cases where the analysis of the difficulty (by means of the Noether machine) also makes the remedy obvious: what needs to be added is precisely the requirement that the theory's Lagrangian $L$ be such that its superpotential is non-trivial (we again remind the reader that a choice of $L$ also fixes the superpotential). We thus propose the following definition of substantive general covariance in response to the `corner charge' part of the Einstein test:
\begin{quote} \textbf{(Corner-SGC)}
A theory has Corner-SGC when, in addition to BGC, its Lagrangian $L$ has a non-trivial superpotential $U_\xi$.\footnote{NB: The corresponding Noether charge will be conserved if we impose boundary conditions that rule out flux leakage, and the Lagrangian $L$ is consistent with those boundary conditions.}
\end{quote}

It is worth pausing to emphasize that the potency of Noether's second theorem is not that of a purely mathematical result. However, just as the character of a line or a stroke can begin to suggest the animation or dynamism of a figure in painting, so too can mathematics be taken up into the mobilization of physical thought. What Noether's second theorem does here is offer physicists a heightened awareness of what the mathematical medium of field theory (understood through the lens of the variational calculus and Noether's theorems) is capable of when it is pressed into the service of physical representation: our discussion thus far is precisely an illustration of how the second theorem can be so recruited! Through that discussion, we come to see that the substantiveness of general covariance is not just about the symmetry of a Lagrangian, but also about how---through that symmetry---some kinds of Lagrangians determine a non-trivial corner charge algebra. 

We now turn to the `Extending PR' part of the Einstein test, for which (Corner-SGC) is a necessary but insufficient condition, because of the aforementioned potential tension between the Noether charge and the Hamiltonian charge relative to some boundary condition.
Note that for a gauge theory (where we mean to include both General Relativity and Yang-Mills theory), whatever the symmetries of generalized PR are, they will need to be generated by the non-trivial corner charges from (Corner-SGC). 
However, there is an added complication present in generalizing Embodied PR: we are now considering the symmetries of dynamically isolated subsystems, which are generated by the Hamiltonian charges for the boundary conditions that model dynamical isolation---thus, we will only succeed in extending PR if the corner charges coming from the trivial Noether current of some Lagrangian are also Hamiltonian charges. In Section 3.2.2, we saw an example (Maxwell gauge theory with Dirichlet boundary conditions) where this coincidence clearly obtains: we can think of it as a regularized (finite) version of a case in which the asymptotically rigid $U(1)$ symmetries of some subsystem provide a generalized version of PR relative to an environment subsystem (where the environment has been idealized away into the boundary conditions). 
However, as we are about to see in Section 3.4, the corner charges from some particular Lagrangian can also fail to be Hamiltonian charges (relative to some choice of dynamically isolated boundary conditions).
We thus propose the following more demanding notion of substantive general covariance in response to the `extending PR' part of the Einstein test:
\begin{quote}
    \textbf{(PR-SGC)} Fix boundary conditions that represent `dynamical isolation' in the context of some theory. The theory's Lagrangian $L$ has PR-SGC just in case, in addition to satisfying Corner-SGC, the Noether charge coming from $L$ is also a Hamiltonian charge relative to the `dynamically isolated' boundary conditions. 
\end{quote}
We now turn to a brief consideration of General Relativity, which illustrates the subtleties of PR-SGC. 

\subsection{Revisiting the case of General Relativity}

In \cite{freidel2021extended}, the general way in which Noether charges can fail to be Hamiltonian charges is considered, and this framework is then applied to the case of General Relativity (Section 3.4 of \cite{freidel2021extended}). Here we only summarize the relevant features for us to revisit the Einstein-Klein dispute. 

First, consider the Einstein-Hilbert Lagrangian $L_{EH}$. We can follow through the covariant phase space recipe sketched in Section 3.1 and compute the superpotential corresponding to $L_{EH}$, which turns out to yield what is known as the Komar charge (see (2.38) of \cite{freidel2020edge} for a definition). 
This shows that we have satisfied (Corner-SGC).

Next, let us consider whether this Lagrangian satisfies (PR-SGC).
To do so, suppose that we wish to use Dirichlet boundary conditions to model an isolated system (or a finite boundary regularization thereof): that is to say, we wish to set the induced metric $\bar{g}_{ab}$ on the timelike boundary to a fixed value, thereby also setting $\delta \bar{g}_{ab} \stackrel{\partial M}{=} 0$. More specifically, we typically take $\bar{g}_{ab}$ to be the Minkowski metric, thus leading to an `asymptotically flat' solution in the limit where we take the boundary to infinity. 
It is well-known that given these Dirichlet boundary conditions, one can construct a Hamiltonian charge; however, this is not the Komar charge, but a different quasi-local charge known as the Brown-York charge (see (2.31) of \cite{freidel2020edge} for a definition). 

Thus we have the complication that we discussed earlier: if one chooses $L_{EH}$ as one's Lagrangian and the Dirichlet boundary condition as the `isolated boundary condition' in the `Extending PR' part of the Einstein test, then one ends up with a mismatch between the Noether charge (the Komar charge) and the the Hamiltonian charge (the Brown-York charge).
The resolution of this complication, discussed in \cite{harlow2020covariant, freidel2021extended}, is to choose a different Lagrangian---which is associated with a different boundary condition---in order to satisfy (PR-SGC). In particular, if we choose the Lagrangian $L = L_{EH} + d\ell_{GHY}$, where $\ell_{GHY}$ is the celebrated Gibbons-Hawking-York boundary term, then the corresponding Noether charge is precisely the Brown-York charge, so we have now brought our Lagrangian into alignment with the boundary condition---and corner charge---that we are after. Furthermore, as \cite{harlow2020covariant} shows, if we assume that the metric tends towards the Minkowski metric in the asymptotic limit, then we recover precisely the ADM charges for an asymptotically flat spacetime, which is precisely what one expects of an isolated subsystem in General Relativity. In particular, we note that these charges generate the `asymptotic Poincare symmetries', which are the generalized analogs of the rigid Poincare symmetries (from the standard PR) in this context.\footnote{Here we are only discussing spatial infinity; once one takes into account null infinity, one is forced to consider the infinite-dimensional BMS group if one wants a rich set of solutions.} We thus see that $L = L_{EH} + d\ell_{GHY}$ satisfies (PR-SGC), but $L_{EH}$ does not.

With these clarifications in hand, we can also return to an assessment of the dispute between Einstein and Klein that we mentioned in our Introduction (we refer the reader to \cite{de2021noether} for a complementary analysis of this dispute, with an emphasis on holographic renormalization). 
Is it really true that Einstein was terribly confused about the status of conserved charges in General Relativity, and that Noether's Second Theorem served to highlight his confusion by securing Klein's claim that General Relativity does not have any physically meaninful conservation laws?
To answer this question, let us consider several key points from Einstein's March 1918 response to Klein (letter 480 in \cite{EinsteinVol8}):
\begin{itemize}
    \item Einstein says the importance of the quantity $J_\xi= \mathfrak{T} + \mathfrak{t}$ (the on-shell exact Noether current, cf. (8) in Section 3.1) is its physical interpretation as the energy of a point mass when we go very far away from the subsystem; in other words, when we treat the subsystem as dynamically isolated from an environment.
    \item He also says that the possibility of this physical interpretation is underwritten by the fact that $J_\xi$ (on-shell) is an \textit{exact} term, presumably because he knows that via Stokes' theorem, a bulk integral of an exact term can be transformed into a boundary integral (which is why we call such charges `quasi-local'). 
    \item Einstein emphasizes that the equations of motion are necessary in order to show that $J_\xi$ is exact, and thus $J_\xi$ should be regarded as physically meaningful.
\end{itemize}
And then, a litle over a week later, in letter 492 to Klein \cite{EinsteinVol8}, Einstein goes on to place a heuristic `isolated' boundary condition on the subsystem that allows for the construction of a conserved charge!

We can now ask ourselves how Einstein's more physically-oriented (and more mathematically imprecise) strategy looks with the clarity of hindsight. 
It would seem that Einstein already understood much of the intuitive physical content that can be expressed by means of Noether's Second Theorem: certainly, he understood that the \textit{distinctive} qualitative feature of a gauge theory (implied by Noether's Second Theorem) is the on-shell exactness of the Noether current, and that this feature leads to a boundary charge; and he was undoubtedly sensitive to the need to \textit{choose} boundary conditions in order to construct such a charge.
In all this, we believe that Einstein was vindicated in resisting Klein's criticisms.
On the other hand, the place where Einstein's understanding was most lacking was in his grasp of (i) the relationship between a choice of $L$ and a choice of boundary conditions, (ii) the potential mismatch between the Noether charge corresponding to some $L$ and the Hamiltonian charge corresponding to some choice of boundary conditions, and (iii) the potential appearance of infinite-dimensional symmetries (such as the BMS group at null infinity) in the asymptotic limit. As we discussed in Section 3.2, this lack is closely related to his struggle to find an appropriate extension of the Relativity Principle to the context of General Relativity. 

\subsection{Background independence}

While this is not the place for us to offer a fully-developed account of `background independence', it seems appropriate to make some brief remarks about how this notion is related to our preferred conceptualization of substantive general covariance, especially since Pooley \cite{pooley2017background} has attempted to pour cold water on attempts by physicists \cite{rovelli2007quantum, smolin2006case, rovelli2004quantum} to find a connection between \textit{some} form of general covariance and background independence.\footnote{According to Pooley, Smolin \cite{smolin2006case} seems to identify background independence with BGC, whereas Rovelli \cite{rovelli2007quantum} seems to want a slightly more full-blooded form of general covariance.} 

Roughly speaking, Pooley's narrative runs as follows. 
There is a standard line of thought in the quantum gravity literature according to which a theory's `general covariance' is tightly linked to its background independence, conceived of as a theory's spacetime structure being dynamical, as opposed to playing the role of a `fixed background'.
As evidence, Pooley cites Rovelli's claim \cite{rovelli2004quantum} that the background independence of classical GR is `...realized by the gauge-invariance of the action under active diffeomorphisms.' 
But, Pooley goes on to argue, this and similar claims seem to run counter to the moral of Kretschmann's point that any theory can be formulated in a generally covariant manner---and it is so much the worse for these claims!
Pooley goes on to prosecute his case in various ways, but the part of his argument that will be of interest to us here is his attempt to show that an identification of background independence with (what we call) BGC misclassifies certain theories which are intuitively background dependent as background independent.

There are two cases of relevance for us.
First, Pooley says that the theory (that we call) \textbf{SR2} clearly satisfies BGC, but seems to be the same theory as \textbf{SR1}, which is a paradigm case of background dependence.
Second, Pooley introduces a theory that we shall call \textbf{SR3}, defined by the Lagrangian\footnote{This theory was introduced by Rosen and rediscovered by Sorkin (see \cite{pooley2017background} for the history and references). Its most careful analysis so far seems to have been by \cite{giulini2007remarks}.} 
\begin{equation}
    L^{\text{\textbf{SR3}}} = \sqrt{-g} ( -\frac{1}{2} \nabla_a \phi \nabla^a \phi + \frac{1}{4} \lambda^{abcd} R_{abcd} ).
\end{equation}
Pooley again notes that this theory satisfies BGC, but seems to be background dependent, and so the identification of BGC with background independence seems to misclassify the example.
On this basis, Pooley concludes that `...one should recognise that we are now far past the point where one might hope to articulate a simple and illuminating connection between diffeomorphism invariance and background independence.'

We of course agree with Pooley that \textbf{SR2} and \textbf{SR3} are clear cases of Lagrangians that satisfy BGC. But as we have just argued above, BGC is not plausibly a \textit{substantive} notion of general covariance (certainly, it would not have been to Einstein in light of his desiderata), and the Noetherian analysis that reveals its implausibility also suggests a minimal notion of substantive general covariance, viz. what we called Corner-SGC in Section 3.3; thus, it would be premature to judge on the basis of BGC that there is no illuminating connection between some form of general covariance and background independence.
With this background in mind, we now offer a conjecture for a simple and illuminating connection between background independence and Corner-SGC:
\begin{quote}
    \textbf{Conjecture:} A theory is background independent only if its Lagrangian satisfies Corner-SGC with respect to spacetime diffeomorphisms.
\end{quote}
The intuition driving this conjecture is that a non-trivial corner charge associated with diffeomorphisms (and the Lagrangian) is required for spacetime structure to have \textit{dynamical} degrees of freedom, and thus for the theory to be background independent. 

How does Pooley's misclassification argument against BGC---using \textbf{SR2} and \textbf{SR3} as test cases---fare against our conjecture?
Let us take the test cases in turn. 
The case of \textbf{SR2} is especially simple: as we discussed in Section 3.2.1, this theory does not satisfy (Corner-SGC) because its Noether current is identically zero and its corner charge is trivial.
Thus, according to our conjecture, this theory counts as background dependent, and we see that Pooley's argument is no threat to the conjecture.

The case of \textbf{SR3} is somewhat more interesting. 
When we put this theory through the Noether machine, we find that the equations of motion are $g_{ab} \nabla^a \nabla^b \phi = 0$, $R_{abcd} (g) = 0$, and $\nabla_a \nabla_d (\lambda^{abcd} + \lambda^{acbd}) = T^{bc}$. We also obtain a set of non-trivial Noether identities and a pre-symplectic potential current
\begin{equation}
    \theta = - \partial^a \phi \delta \phi + \frac{1}{2} (\nabla_d \lambda^{abcd} \delta g_{bc} - \lambda^{abcd} \nabla_{d} \delta g_{bc} )
\end{equation}
from which we can compute the Noether current
\begin{IEEEeqnarray}{rCl}
    J_{\xi}^a &=& - T^{a}_b \xi^b + \frac{1}{2} \nabla_b (\nabla_d \lambda^{a (bc)d} \xi_c) 
    - \frac{1}{2} \nabla_d ( \lambda^{a(bc)d} \nabla_b \xi_c ) \nonumber\\
&& -\> \lambda^{abcd} \nabla_d \nabla_b \xi_c +\frac{1}{4}( \nabla_d \lambda^{abcd} \nabla_c \xi _b - \lambda^{abcd} \nabla_d \nabla_c \xi _b).
\end{IEEEeqnarray}
From this we can compute the superpotential $2$-form $U_{\xi} = \epsilon_{ab} \lambda^{abcd} \nabla_c \xi_d$, where $\epsilon_{ab}$ is an area $2$-form obtained by contracting $\partial_a$ and $\partial_b$ with the volume form. In other words, we see that this theory has a non-trivial corner Noether charge, and thus our conjecture classifies it as potentially background independent.

What then are we to make of Pooley's `intuition' that \textbf{SR3} is background dependent? 
Note that the basis for this intuition seems to have two parts. The first is the uncontroversial observation that \textbf{SR3}'s equations of motion $R^{abcd}(g)=0$ can be replicated in a non-Lagrangian theory, viz. the theory specified just by the equations of motion $\nabla_a \nabla^a \phi =0$ and $R^{a}_{bcd}=0$. 
The second is Pooley's claim that if the latter theory is regarded as background dependent, then the former theory should also be regarded as background dependent.

It will help to introduce an analogy to put the observation in context and to expose the defects of Pooley's line of thought here.
Consider an example in which we couple some matter field degrees of freedom to a $U(1)$ BF theory, which in 4 dimensions has the Lagrangian $B \wedge dA$, where the dynamical fields are a $2$-form $B$ and a gauge field $A$ with curvature $F=dA$; the equations of motion of the pure BF theory are $F=0$ and $dB=0$.
We will take this example to be our first theory, to be compared to a second theory in which the same matter field degrees of freedom are set against the fixed background of a flat $U(1)$ connection, which is of course described by the equation $F=0$; in the former case, we can think of the flatness condition as being enforced by a Lagrange multiplier field, whereas it is imposed by fiat in the latter case.
If we applied Pooley's reasoning to this analogous scenario, then we would say that since the latter theory is background dependent (now in the gauge field, and not the spacetime, sense), we should also regard the former theory as background dependent. 
But now it should be evident that this line of thought cannot be right, because a BF theory is a topological field theory, which is supposed to be one of the paradigm cases of background independence!
The mistake here is to conflate local dynamical degrees of freedom---which both theories lack---with global dynamical degrees of freedom, which only BF theory possesses, and in virtue of which it is called `topological'. 
For instance, in the case of BF (or Chern-Simons) theories of gravity, what we would standardly say is that the spacetime is dynamical in the sense that it has these global degrees of freedom.

With the analogy in hand, we are now in a position to better understand the physics of \textbf{SR3}. It is precisely a BF-like theory, and as noted by Giulini \cite{giulini2007remarks}, the spacetime part has global---but not local---dynamical degrees of freedom which make the solution space for \textbf{SR3} inequivalent to the theory specified just by the equations of motion $\nabla_a \nabla^a \phi =0$ and $R^{a}_{bcd}=0$. 
The equation of motion for $\lambda$, in which this dynamical parameter now appears as a potential for the matter field's stress-energy, and the non-trivial corner charge from $U_\xi$, serve to describe how the global spacetime dynamics is coupled to the dynamics of the matter field. 

To sum up: we again see that in the case of \textbf{SR3}, Pooley's misclassification argument poses no threat to our conjecture; on the contrary, a careful Noetherian analysis of \textbf{SR3} indicates that Pooley has misinterpreted this theory.

\section{Conclusion}

Noether's theorems, as we said early on in the Introduction, do not in themselves resolve any questions about the interpretation of General Relativity, or lend clarity to the \textit{physical} (not merely mathematical) dispute between Einstein and Klein. Their true glory lies in their being a uniquely subtle medium for physical representation---a set of mathematical materials that physicists can bend to the ends of a certain kind of representational activity. 

We pay tribute to the author of these theorems by closing with a broader philosophical reflection of the kind of activity into which her theorems are taken up. When describing the productive activity of house-building, Aristotle introduces the idea of \textit{dunamis}, i.e. the potency or capacity of the building materials, which the builder actualizes in pursuing his goal of building a house. This kind of activity Aristotle calls a \textit{kinesis}, because it aims at a further product outside itself, namely the completed house. And he goes on to contrast \textit{kinesis} with what he calls \textit{energeia}, i.e. the kind of activity which aims at nothing beyond itself and is its own end (two of his examples are `seeing' and `dancing'). 

Evidently, Noether's theorems are a \textit{dunamis} in Aristotle's sense: they have an incredible capacity to be actualized in representation, which Einstein merely intuited, and which we have tried to flesh out more fully in Section 3, although we still only managed to scratch the surface. 
A further interesting question is whether this activity into which Noether's theorems are taken up---the activity of using them to represent the material world---is more akin to \textit{kinesis} or \textit{energeia}. And here we would like to make the suggestion that \textit{energeia} is the better description, for the kind of insight achieved by absorbing physical subject matter into the framework of Noether's theorems (e.g. in the reconceptualization of Section 3.2) leads to ever new insights that energize and propel the physicist's theorizing further forward---the end here is internal to the activity of doing theoretical physics, and not a further product that is external to the activity.   
And perhaps this is close to what Noether had in mind when she said that `My methods are really methods of working and thinking; this is why they have crept in everywhere anonymously.'

\section*{Acknowledgements}
We thanks James Read for helpful comments on this draft. NT's research is partially supported by JTF grant 61521 and NSF grant 1947155. LF's research at Perimeter Institute is supported in part by the Government of Canada through the Department of Innovation, Science and Industry Canada and by the Province of Ontario through the Ministry of Colleges and Universities. 
\newpage
\printbibliography 

\end{document}